\RequirePackage{amsmath} 
\documentclass[12pt]{iopart}

\usepackage{graphicx}%

\usepackage{cite}
\usepackage{amsfonts}
\usepackage{braket}
\usepackage{tikz}
\usetikzlibrary{angles,quotes}
\usetikzlibrary{fit,positioning}
\usetikzlibrary{arrows, automata}

\newtheorem{theorem}{Theorem}
\newtheorem{proof}{Proof}

\newtheorem{remark}{Remark}
\newtheorem{definition}{Definition}
\begin{document}

\title[Entanglement in Directed Graph States]{Entanglement in Directed Graph States}

\author{Lucio De Simone and Roberto Franzosi}

\address{Department of Physical Sciences, Earth and Environment, University of Siena, Via Roma 56, 53100 Siena, Italy}
\address{INFN Sezione di Perugia, INFN, 06123 Perugia, Italy}

\ead{l.desimone3@student.unisi.it and roberto.franzosi@unisi.it}
\vspace{10pt}
\begin{indented}
\item[]May 2025
\end{indented}

\begin{abstract}
We investigate a family of quantum states defined by directed graphs, where the oriented edges represent interactions between ordered qubits. As a measure of entanglement, we adopt the Entanglement Distance—a quantity derived from the Fubini–Study metric on the system’s projective Hilbert space.

We demonstrate that this measure is entirely determined by the vertex degree distribution and remains invariant under vertex relabeling, underscoring its topological nature. Consequently, the entanglement depends solely on the total degree of each vertex, making it insensitive to the distinction between incoming and outgoing edges.

These findings offer a geometric interpretation of quantum correlations and entanglement in complex systems, with promising implications for the design and analysis of quantum networks.
\end{abstract}

%
%
%
%
%

\section{Introduction}\label{sec1}
Let $\mathcal{H}=\mathcal{H}^{\otimes M}_2$ be the Hilbert space of a system of $M$ qubits. 
Let $\mathcal{PH}$ denote the corresponding projective Hilbert space, i.e., the set of equivalence classes of non-zero vectors
$\ket{\psi}\in \mathcal{H}$, under the relation $\sim$ defined as
$\ket{\psi} \sim \ket{\phi}$ if and only if $\ket{\psi} =\alpha \ket{\phi}$, for some $\alpha \in \mathbb{C}$, $\alpha \neq 0$.
The Fubini-Study metric 
 provides the infinitesimal distance between two neighbouring states $\ket{\psi}$ $\ket{\psi} + \ket{d\psi}$ in $\mathcal{H}$, and is given by
 \cite{cmp/1103908308,gibbons},
\begin{equation}
    d_{FS}(\ket{\psi} + \ket{d\psi},\ket{\psi}) = 
    \left[
    \braket{d\psi|d\psi} -|\braket{\psi|d\psi}|^2
    \right]^{1/2} \, .
    \label{dFS}
\end{equation}
From this expression, one obtains the metric tensor
\begin{equation}
g_{\mu \nu} = \langle \partial_\mu \psi|\partial_\nu \psi \rangle -
\langle \partial_\mu \psi| \psi \rangle \langle \psi|\partial_\nu \psi \rangle 
\label{mtFS}
\end{equation}
 which serves as the foundation for defining the Entanglement Distance (ED) \cite{cocchiarella_entanglement_2020, vesperini_entanglement_2023,Vesperini_2024}.
 
In the case of a system of $M$ qubits, the Entanglement Distance per qubit is given by \cite{nourmandipour_entanglement_2021, vesperini_correlations_2023, vafafard_multipartite_2022}
\begin{equation}\label{ent_dist}
E(\ket{\psi})=1-\dfrac{1}{M}\sum_{i=1}^{M}||\bra{\psi}\boldsymbol{\sigma}^{(i)}\ket{\psi}||^2 \, ,
\end{equation}
where $\boldsymbol{\sigma}^{(i)}=(\sigma^{(i)}_x,\sigma^{(i)}_y,\sigma^{(i)}_z)$ is the vector of the Pauli matrices acting on the $i$th qubit. 

\begin{remark}
The ED per qubit equals $1$ ($E(\ket{\psi})=1$) if $\ket{\psi}$ is maximally entangled, and vanishes ($E(\ket{\psi})=0$) when the state is fully separable.   
\end{remark}
We adopt Eq. \eqref{ent_dist} as the entanglement measure for a general pure quantum state of multiple qubits.
In the present study, we examine states associated with directed graphs \cite{hein_multi-party_2004, hein2006entanglement, gnatenkoGeometricMeasureEntanglement2022, PhysRevA.105.042418, PhysRevA.78.042309}, constructed according to the following definitions.
\begin{definition}[Graph State\label{def1}]
Let $G(V,L)$ be a directed graph, where $V$ is the set of vertices, $V=\{1,\ldots,M\}$, $M\in \mathbb{N}^+$, and $L$ is the set of ordered couples of elements of $V$, identifying the set of oriented edges, $L=\{(a,b)|\ a,b\in V\}$. Let $\ket{\phi}^a$ be the state associated with the vertex $a\in V$ and let $U_{ab}$ be a nonlocal operator acting on the subspaces of the vertices $a$ and $b$, with $a,b\in V$. We define the graph state corresponding to $G(V,L)$ as 
\begin{equation}
    \ket{G}=\prod_{(a,b) \in L}{U}_{ab}\ket{\phi}^{\otimes M} \, .
\end{equation}
\end{definition}
\begin{remark}
    According to this definition, each vertex $a\in V$ is associated with the state $\ket{\phi}^a$. In particular, the graph state corresponding to the empty graph, i.e., a directed graph $G(V,L)$ with $L = \emptyset$, is given by the product state $\ket{\phi}^{\otimes M}$.
\end{remark}

\begin{definition}[Ordering-free graph state\label{def2}]
    An ordering-free graph state is a graph state in which there is no ordering of the edges in $G(V,L)$, i.e. all two-particle unitary operators $U_{ab}$ commutate, that is $[U_{ab},U_{bc}]=0$, $\forall a,b,c\in V$.
\end{definition}
We further assume that the interaction operator $U_{ab}$ is the same for all the edges in $G(V,L)$. For a general controlled-$\bar{U}$ operator $U_{ab}=\Pi_0^{(a)}\mathbb{I}^{(b)}+\Pi_1^{(a)}\bar{U}^{(b)}$, these conditions can be satisfied by restricting the operator $U_{ab}$ to the form
\begin{equation}
\label{oper}
U_{ab}=\Pi_0^{(a)}\mathbb{I}^{(b)}+\Pi_1^{(a)}\bar{U}^{(b)}\,, \hspace{5mm}\bar{U}^{(b)} = e^{-i\psi}
\begin{pmatrix}
e^{i\theta} & 0  \\
0 & {e^{-i\theta}}
\end{pmatrix}\,.
\end{equation}

To generate a maximally entangled state, denoting by $\rho_{\gamma}$, with $\gamma=a,b$, the reduced density operator of a quantum graph state with two qubits and one edge $\ket{G}=U_{ab}\ket{\phi,\phi}$, we uniquely choose the initial state $\ket{\phi}=\alpha_0\ket{0}+\alpha_1\ket{1}$, with $|\alpha_0|^2+|\alpha_1|^2=1$, such that the following conditions are satisfied: \emph{i)} the Hilbert-Schmidt distance $D_{HS}(\rho_{\gamma},\mathbb{I}/2)=\sqrt{\frac{1}{2}\mathrm{tr}\big[(\rho_{\gamma}-\mathbb{I}/2)^{\dagger}(\rho_{\gamma}-\mathbb{I}/2)\big]}$ is minimized; \emph{ii)} the von-Neumann entropy $S(\rho_{\gamma})=-\mathrm{tr}[\rho_\gamma \ln \rho_\gamma]$ is maximized; \emph{iii)} the ED per qubit \eqref{ent_dist} is maximized. Under these conditions, the maximally
entangled state is achieved when $|\alpha_0|=|\alpha_1|=1/\sqrt{2}$. Therefore, from now on, we will adopt the initial
state $\ket{\phi}=(\ket{0} + \ket{1})/\sqrt{2}$ and the operator in Eq. \eqref{oper}.

\section{Entanglement in General Graph Configurations}
\begin{theorem}
 Let $\ket{G}$ be the graph state associated with the ordering-free graph $G(V,L)$. Let the unitary operator $U_{ab}$ be given in a controlled-$\bar{U}$ form as in Eq. \eqref{oper}.
Then, the Entanglement Distance per qubit \eqref{ent_dist} is given by
\begin{equation}
\label{final}
    E(\theta;\{d(i)\})=1-\dfrac{1}{M}\sum_{i\in V} [\cos ({\theta})]^{2d(i)} \, .
\end{equation}
where $E(\theta;\{d(i)\}):=E(\ket{G})$ and $d(i)$ is the degree of the $i$-th qubit.
\end{theorem}
\begin{remark}
    From Eq.\eqref{final}, we note that the entanglement is solely determined by the vertex degree distribution. This invariance underscores the topological nature of the measure applied to quantum networks, highlighting that the structural properties of the graph, rather than the specific orientations of the edges, govern the entanglement distribution. 
\end{remark}
\begin{proof}
It is sufficient to consider the quantity
\begin{equation}
\label{single-qubit_ED}
    E^{(i)}(\ket{G}) = 1 - ||\bra{G}\boldsymbol{\sigma}^{(i)}\ket{G}||^2\,,
\end{equation}
that is, the contribution of the $i$-th vertex to the entanglement distance. Let $\Gamma_\rightarrow(i)$ ($\Gamma_\leftarrow(i)$) denote the set of vertex labels connected to $i$ by an outgoing (incoming) link, 
\begin{align}
\Gamma_\rightarrow(i) = \{j\in V| (i,j)\in L  \} \, , \\
\Gamma_\leftarrow(i) = \{j\in V|  (j,i) \in L \} \, .
\end{align}
The set $\Gamma(i)=\Gamma_{\rightarrow}(i)\cup\Gamma_{\leftarrow}(i)$ thus is the set of vertices connected to $i$ by an edge, and $d(i):=|\Gamma(i)|$ the degree of the vertex $i$. We explicitly make a clear distinction between outgoing and incoming links to ultimately show that this difference does not affect the entanglement. The proof is organized in three steps, that is, when \emph{i)} $\Gamma_\leftarrow(i)=\emptyset$, \emph{ii)} $\Gamma_\rightarrow(i)=\emptyset$ and \emph{iii)} $\Gamma_\leftarrow(i)\neq\emptyset$ and $\Gamma_\rightarrow(i)\neq\emptyset$.

\emph{i)} Let us consider now the case where the vertex $i$ is connected
to $d_{\rightarrow}(i)$ vertices, $j_1,\ldots,j_{d_{\rightarrow}(i)}\in \Gamma_\rightarrow(i)$, by $d(i)$ outgoing links.
Fig. (\ref{fig5}) shows an example of this case.
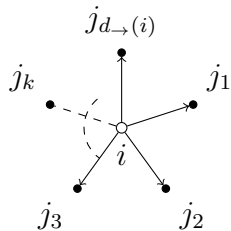
\begin{figure}
\begin{center}
\begin{tikzpicture}
\coordinate (a) at (0,0);
\coordinate (b) at ({cos(234)}, {sin(234)});
\coordinate (ax) at ({cos(162)}, {sin(162)});
\coordinate (ax) at ({cos(130)}, {sin(130)});
  \node[draw, fill=white, circle, inner sep=1pt, label=below:$i$,minimum size=4pt] (1) at (0, 0) {};
  \node[draw, fill=black, circle, inner sep=1pt, label=above:$j_{d_{\rightarrow}(i)}$] (2) at ({cos(90)}, {sin(90)}) {}; 
  \node[draw, fill=black, circle, inner sep=1pt, label=above right:$j_1$] (3) at ({cos(18)}, {sin(18)}) {};  
  \node[draw, fill=black, circle, inner sep=1pt, label=below right:$j_2$] (4) at ({cos(306)}, {sin(306)}) {};
  \node[draw, fill=black, circle, inner sep=1pt, label=below left:$j_3$] (5) at ({cos(234)}, {sin(234)}) {}; 
  \node[draw, fill=black, circle, inner sep=1pt, label=above left:$j_k$] (6) at ({cos(162)}, {sin(162)}) {};  

  \path[->] (1) edge node {} (2);
  \path[->] (1) edge node {} (3);
  \path[->] (1) edge node {} (4);
  \path[->] (1) edge node {} (5);
  \path[dashed] (1) edge node {} (6);
  \pic[draw, dashed, angle radius=0.5cm,angle eccentricity=0] {angle=ax--a--b};
  \pic[draw, dashed, angle radius=0.5cm,angle eccentricity=0] {angle=b--a--b};
\end{tikzpicture}
\caption{This is an example with $\Gamma_\rightarrow(i)=\{j_1,\ldots,j_{d_{\rightarrow}(i)}\}$, $\Gamma_\leftarrow(i)=\emptyset$, and $U_{tot}=\prod^{d_{\rightarrow}(i)}_{k=1} {U}_{ij_k}$.\label{fig5}}
\end{center}
\end{figure}
In this case $\Gamma_\leftarrow(i)=\emptyset$, and the index $i$ of the Pauli operators $\boldsymbol{\sigma}^{(i)}$ in expression \eqref{single-qubit_ED} always appears as the first index of each operator ${U}_{ij_k}$, $k=1,\ldots,d_{\rightarrow}(i)$.
The full unitary operator results
\begin{equation}
\label{1}
{U}_{tot}=\! \! \prod_{j\in\Gamma_\rightarrow(i)}{U}_{ij}=\Pi^{(i)}_0\!\! \prod_{j\in\Gamma_\rightarrow(i)}\mathbb{I}^{(j)}+\Pi^{(i)}_1\prod_{j\in\Gamma_\rightarrow(i)}\bar{U}^{(j)} \, ,
\end{equation}
and, using the properties of Pauli matrices, one obtains
\begin{equation}
{U}^{\dagger}_{tot}\sigma^{(i)}_{x,y} {U}_{tot}=\Pi^{(i)}_0\sigma^{(i)}_{x,y} \! \! \prod_{j\in\Gamma_\rightarrow(i)} \! \!\bar{U}^{(j)}+\sigma^{(i)}_{x,y}\Pi^{(i)}_0 \! \! \prod_{j\in\Gamma_\rightarrow(i)} \! \! \bar{U}^{(j)\dagger} \, ,
\end{equation}
and
\begin{equation}
{U}^{\dagger}_{tot}\sigma^{(i)}_z {U}_{tot}=\sigma^{(i)}_z\prod_{j\in\Gamma_\rightarrow(i)}\mathbb{I}^{(j)} \, .
\end{equation}
We have
\begin{eqnarray}
   \bra{\phi}^{\otimes M}&
    {U}^{\dagger}_{tot}\boldsymbol{\sigma}^{(i)}{U}_{tot}
    \ket{\phi}^{\otimes M} = \nonumber \\ \nonumber
    =&\dfrac{1}{2}\left(\bra{\phi}\bar{U}\ket{\phi}^{d_{\rightarrow}(i)}+\text{c.c},\,i\bra{\phi}\bar{U}^{\dagger}\ket{\phi}^{d_{\rightarrow}(i)}+\text{c.c},\,0\right)=\\ 
    &=\cos^{d_{\rightarrow}(i)}(\theta)\big(\cos(d_{\rightarrow}(i)\psi),-\sin(d_{\rightarrow}(i)\psi),0\big)\, .
\end{eqnarray}
Through direct calculations, we find that the contribution of vertex $i$ to the ED is
\begin{equation}
E^{(i)} =1 - [\cos ({\theta})]^{2d_{\rightarrow}(i)} \, .
\end{equation}

\emph{ii)} Let us consider now the case where the vertex $i$ is connected
to $d_{\leftarrow}(i)$ vertices, $j_1,\ldots,j_{d_{\leftarrow}(i)}\in \Gamma_\leftarrow(i)$, by $d_{\leftarrow}(i)$ incoming links. Fig. (\ref{fig6}) shows an example of this case.
In this case $\Gamma_\rightarrow(i)=\emptyset$, and the index $i$ of the Pauli operators $\boldsymbol{\sigma}^{(i)}$ in expression \eqref{single-qubit_ED} always appears as the second index of each operator ${U}_{j_ki}$, $k=1,\dots,d_{\leftarrow}(i)$.
We have
\begin{equation}
    U_{tot} = \prod_{j\in\Gamma_\leftarrow(i)}{U}_{ji}=
    \prod_{j\in\Gamma_\leftarrow(i)} \left( \mathbb{I}^{(i)}\Pi^{(j)}_0 +\bar{U}^{(i)}\Pi^{(j)}_1
    \right)\, .
\end{equation}
\begin{figure}
\begin{center}
\begin{tikzpicture}
\coordinate (a) at (0,0);
\coordinate (b) at ({cos(234)}, {sin(234)});
\coordinate (ax) at ({cos(162)}, {sin(162)});
\coordinate (ax) at ({cos(130)}, {sin(130)});
  \node[draw, fill=white, circle, inner sep=1pt, label=below:$i$,minimum size=4pt] (1) at (0, 0) {};
  \node[draw, fill=black, circle, inner sep=1pt, label=above:$j_{d_{\leftarrow}(i)}$] (2) at ({cos(90)}, {sin(90)}) {}; 
  \node[draw, fill=black, circle, inner sep=1pt, label=above right:$j_1$] (3) at ({cos(18)}, {sin(18)}) {};  
  \node[draw, fill=black, circle, inner sep=1pt, label=below right:$j_2$] (4) at ({cos(306)}, {sin(306)}) {};
  \node[draw, fill=black, circle, inner sep=1pt, label=below left:$j_3$] (5) at ({cos(234)}, {sin(234)}) {}; 
  \node[draw, fill=black, circle, inner sep=1pt, label=above left:$j_k$] (6) at ({cos(162)}, {sin(162)}) {};  

  \path[<-] (1) edge node {} (2);
  \path[<-] (1) edge node {} (3);
  \path[<-] (1) edge node {} (4);
  \path[<-] (1) edge node {} (5);
  \path[dashed] (1) edge node {} (6);
  \pic[draw, dashed, angle radius=0.5cm,angle eccentricity=0] {angle=ax--a--b};
  \pic[draw, dashed, angle radius=0.5cm,angle eccentricity=0] {angle=b--a--b};
\end{tikzpicture}
\caption{This is an example with $\Gamma_{\leftarrow}(i)=\{j_1,\ldots,j_{d_{\leftarrow}(i)}\}$ and $U_{tot}=\prod^{d_{\leftarrow}(i)}_{k=1} {U}_{j_k i}$.\label{fig6}}
\end{center}
\end{figure}
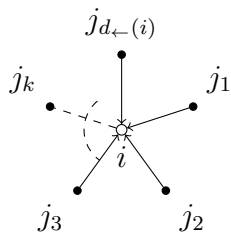
By induction, we prove that this operator has the general form
\begin{equation}
\label{2}
    U_{tot}^{(n)}=\sum_{k=0}^{n} P^{(n)}_k \bar{U}^{(i)^{k}} \,
\end{equation}
where $n=d_{\leftarrow}(i)$ and $P^{(n)}_k$ is the symmetric sum over all the permutations of the tensor product consisting of $n-k$ projectors $\Pi^{j_m}_0$ and $k$ projectors $\Pi^{j_m}_1$, with $m = 1,\dots,d_{\leftarrow}(i)$, with the operators $P^{(n)}_k$ satisfying the orthogonality relation $P_k^{(n)}P_{k'}^{(n)}=P_k^{(n)}\delta_{k,k'}$. In fact, for $n=1$, we have $U_{tot}^{(1)}=U_{j_1 i}$. For $n+1$, the operator $U_{tot}^{(n+1)}=U_{j_{n+1} i} U_{tot}^{(n)}$ can be expanded as
\begin{eqnarray}
U_{tot}^{(n+1)}=\,&\Pi_0^{(j_{n+1})}P_0^{(n)}\mathbb{I}^{(i)}+\Pi_1^{(j_{n+1})}P_n^{(n)}\bar{U}^{(i)^{n+1}}+ \nonumber \\&+\sum_{k=1}^n \big[\Pi_0^{(j_{n+1})}P_k^{(n)}+\Pi_1^{(j_{n+1})}P_{k-1}^{(n)} \big] \bar{U}^{(i)^{k}}\, ,
\end{eqnarray}
where we recognize that the terms in the expression correspond to the symmetric projectors of order $n+1$, i.e. we identify $P_0^{(n+1)}=\Pi_0^{(j_{n+1})}P_0^{(n)}$, $P_{n+1}^{(n+1)}=\Pi_1^{(j_{n+1})}P_n^{(n)}$ and $P_k^{(n+1)}=\Pi_0^{(j_{n+1})}P_k^{(n)}+\Pi_1^{(j_{n+1})}P_{k-1}^{(n)}$. Using the operator (\ref{2}), by direct calculation, one gets
\begin{eqnarray}
        & \bra{\phi}^{\otimes M}
    {U}^{\dagger}_{tot}\boldsymbol{\sigma}^{(i)}{U}_{tot}
    \ket{\phi}^{\otimes M} = \nonumber  \\ \nonumber
        & =\sum_{k=0}^{d(i)}\bra{\phi}^{\otimes M-1}P_k^{\left(d_{\leftarrow}(i)\right)}\ket{\phi}^{\otimes M-1}\bra{\phi}\bar{U}^{\dagger k}\boldsymbol{\sigma}\bar{U}^k\ket{\phi}=\\ 
        &=\cos^{d(i)}(\theta) \big(\cos\left(d_{\leftarrow}(i)\theta\right), -\sin\left(d_{\leftarrow}(i)\theta\right),0\big) \, ,
\end{eqnarray}
from which we obtain again
\begin{equation}
\label{leftPauli}
E^{(i)} =1 - [\cos ({\theta})]^{2d_{\leftarrow}(i)} \, .
\end{equation}

\emph{iii)}
In this last case, we consider the combined configuration of the cases \emph{i)} and \emph{ii)}, namely where the vertex $i$ is connected to $d_{\rightarrow}(i)$ vertices, $j_1,\dots,j_{d_{\rightarrow}(i)}\in\Gamma_{\rightarrow}(i)$, by $d_{\rightarrow}(i)$ outgoing links, and to $d_{\leftarrow}(i)$ vertices, $m_1,\dots,m_{d_{\leftarrow}(i)}\in\Gamma_{\leftarrow}(i)$, by $d_{\leftarrow}(i)$ incoming links. Fig. (\ref{fig7}) shows an example of this case. 
\begin{figure}[h]
\begin{center}
\begin{tikzpicture}
\coordinate (a) at (0,0);
\coordinate (b) at ({cos(0)}, {sin(0)});
\coordinate (bf) at ({cos(270)}, {sin(270)});
\coordinate (c) at ({cos(180)}, {sin(180)});
\coordinate (cf) at ({cos(90)}, {sin(90)});
  \node[draw, fill=white, circle, inner sep=1pt,minimum size=4pt] (1) at (0, 0) {};
  \node at (0.19134, 0.46194) {$i$};
  \node[draw, fill=black, circle, inner sep=1pt, label=below:$j_{d_{\rightarrow}(i)}$] (2) at ({cos(-90)}, {sin(-90)}) {}; 
  \node[draw, fill=black, circle, inner sep=1pt, label=above right:$j_1$] (3) at ({cos(45)}, {sin(45)}) {};  
  \node[draw, fill=black, circle, inner sep=1pt, label=right:$j_2$] (4) at ({cos(0)}, {sin(0)}) {};
  \node[draw, fill=black, circle, inner sep=1pt, label=below right:$j_k$] (5) at ({cos(-45)}, {sin(-45)}) {}; 
  \node[draw, fill=black, circle, inner sep=1pt, label=below left:$m_1$] (6) at ({cos(-135)}, {sin(-135)}) {}; 
  \node[draw, fill=black, circle, inner sep=1pt, label=left:$m_2$] (7) at ({cos(180)}, {sin(180)}) {};
  \node[draw, fill=black, circle, inner sep=1pt, label=above left:$m_p$] (9) at ({cos(135)}, {sin(135)}) {};
  \node[draw, fill=black, circle, inner sep=1pt, label=above:$m_{d_{\leftarrow}(i)}$] (8) at ({cos(90)}, {sin(90)}) {};  

  \path[<-] (1) edge node {} (2);
  \path[<-] (1) edge node {} (3);
  \path[<-] (1) edge node {} (4);
  \path[->] (1) edge node {} (6);
  \path[->] (1) edge node {} (7);
  \path[->] (1) edge node {} (8);
  \path[dashed] (1) edge node {} (5);
  \path[dashed] (1) edge node {} (9);
  \pic[draw, dashed, angle radius=0.5cm,angle eccentricity=0] {angle=bf--a--b};
  \pic[draw, dashed, angle radius=0.5cm,angle eccentricity=0] {angle=cf--a--c};
\end{tikzpicture}
\caption{This is an example with $\Gamma_{\rightarrow}(i)=\{j_1,\ldots,j_{d_{\rightarrow}(i)}\}$, $\Gamma_{\leftarrow}(i)=\{m_1,\ldots,m_{d_{\leftarrow}(i)}\}$ and $U_{tot}=\prod^{d_{\rightarrow}(i)}_{k=1} {U}_{ij_k}\prod^{d_{\leftarrow}(i)}_{p=1}{U}_{m_p i}$.\label{fig7}}
\end{center}
\end{figure}
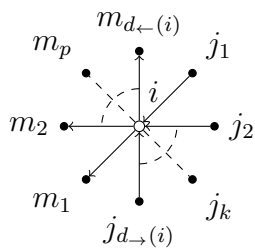
The complete unitary operator results from the product of the operators \eqref{1} and \eqref{2}, with the corresponding adjustments in the indices, and is given by 
\begin{equation}
    U_{tot}=\prod^{d_{\rightarrow}(i)}_{k=1} {U}_{ij_k}\prod^{d_{\leftarrow}(i)}_{p=1}{U}_{m_p i}\, ,
\end{equation}
where, for the following calculations, we denote $U_{\rightarrow}$ and $U_{\leftarrow}$ as $\prod^{d_{\rightarrow}(i)}_{k=1} {U}_{ij_k}$ and $\prod^{d_{\leftarrow}(i)}_{p=1}{U}_{m_p i}$ respectively. For the computation, it is useful to decompose the initial state as $\ket{\phi}^{\otimes M}=\ket{\phi}^{\otimes d_{\rightarrow}(i)}\ket{\phi}^{\otimes d_{\leftarrow}(i)}\ket{\phi}^{(i)}$. In fact, for the expectation value $\bra{\phi}^{d_{\rightarrow}(i)}U_{\rightarrow}^{\dagger}\sigma^{(i)}_{\gamma}U_{\rightarrow}\ket{\phi}^{d_{\rightarrow}(i)}$ one gets
\begin{eqnarray}
     &\cos^{d_{\rightarrow}(i)}(\theta)\left(\ket{0}\bra{1}^{(i)}e^{-id_{\rightarrow}(i)\psi}+\ket{1}\bra{0}^{(i)}e^{id_{\rightarrow}(i)\psi}\right) \nonumber \\
     &\cos^{d_{\rightarrow}(i)}(\theta)\left(-i\ket{0}\bra{1}^{(i)}e^{-id_{\rightarrow}(i)\psi}+i\ket{1}\bra{0}^{(i)}e^{id_{\rightarrow}(i)\psi}\right)
\end{eqnarray}
for $\gamma=x,y$ and $\sigma_z^{(i)}$ for $\gamma=z$. Their expectation value for the state $U_{\leftarrow}\ket{\phi}^{d_{\leftarrow}(i)}$ result in
\begin{eqnarray}
        &\sum_{k=0}^{d_{\leftarrow}(i)}\frac{1}{2^{d_{\leftarrow}(i)}}\binom{d_{\leftarrow}(i)}{k}\cos^{d_{\rightarrow}(i)}(\theta)\left(\ket{0}\bra{1}^{(i)}\,e^{-i(d_{\rightarrow}(i)\psi+2k\theta)}+h.c.\right) \nonumber \\
        &\sum_{k=0}^{d_{\leftarrow}(i)}\frac{1}{2^{d_{\leftarrow}(i)}}\binom{d_{\leftarrow}(i)}{k}\cos^{d_{\rightarrow}(i)}(\theta)\left(-i\ket{0}\bra{1}^{(i)}\,e^{-i(d_{\rightarrow}(i)\psi+2k\theta)}+h.c.\right)\\
        &\sum_{k=0}^{d_{\leftarrow}(i)}\frac{1}{2^{d_{\leftarrow}(i)}}\binom{d_{\leftarrow}(i)}{k}\sigma_z^{(i)}
\end{eqnarray}
and, finally, for the expectation value for $\ket{\phi}^{(i)}$ we have
\begin{eqnarray}
        &\bra{\phi}^{\otimes M}
    {U}^{\dagger}_{tot}\boldsymbol{\sigma}^{(i)}{U}_{tot}
    \ket{\phi}^{\otimes M}= \nonumber \\
    &\cos^{d(i)}(\theta)\big(\cos(d_{\leftarrow}(i)(\psi+\theta)),-\sin(d_{\leftarrow}(i)(\psi+\theta)),0\big)
\end{eqnarray}
from which we obtain again
\begin{equation}
    E^{(i)} =1 - [\cos ({\theta})]^{2d(i)} \, ,
\end{equation}
where $d(i)=d_{\rightarrow}(i)+d_{\leftarrow}(i)$.
\end{proof}

\section{Conclusion}\label{sec13}
We have investigated the entanglement properties of multi-qubit states associated with directed graphs. These states have wide applications in the context of quantum computation and quantum information \cite{nielsen_chuang_2010,nielsen_cluster-state_2006, raussendorf_quantum_2012, PhysRevLett.86.5188}.
We have shown that the entanglement of quantum states defined by directed graphs is fully determined by the graph’s topology—specifically, by the vertex degree distribution—remains invariant under vertex relabeling, and is insensitive to the distinction between incoming and outgoing edges.

\section{Acknowledgements}

We acknowledge the support of the Research Support Plan 2022 – Call for applications for funding allocation to research projects curiosity-driven (F CUR) – Project "Entanglement Protection of Qubits’ Dynamics in a Cavity" – EPQDC and the support from the Italian National Group of Mathematical Physics (GNFM-INdAM). R. F. would like to acknowledge INFN Pisa for the financial support to this activity.

\section*{Declarations}

The data related to the presented results are available.

\bigskip




\bibliography{bibliography}

\end{document}